# Decays of Higgs Boson into Pseudoscalar Meson Pair of Different Flavours

Swee-Ping Chia

*Physics Department, University of Malaya*
*50603 Kuala Lumpur, Malaysia*

**Abstract.** The flavour-changing quark-Higgs coupling does not arise at the tree level in the Standard Model. It is, however, induced by one-loop effects. In this paper, we present an exact calculation of the Higgs-penguin vertex in the 'tHooft-Feynman gauge. Renormalization of the vertex is effected by a prescription by Chia and Chong which gives an expression for the counter term identical to that obtained by employing Ward-Takahashi identity. The on-shell vertex function for the Higgs-penguin vertex is obtained, and its dependence on Higgs mass is investigated. Our calculation is applied to the following process: decays of Higgs boson into pseudoscalar meson pair of different flavours: $K\pi$ and $BK$. The coupling of pseudoscalar meson to quark is effected through a $\gamma_5$ coupling. The decay rates are found to be very small.

## INTRODUCTION

The Higgs boson plays an essential role in the spontaneous breaking of SU(2)xU(1) symmetry in the Standard Model (SM). Its existence is, however, elusive until recently. In July 2012 CERN announced the discovery of a 'Higgs-like' particle by two experimental groups: ATLAS [1] and CMS [2]. The Higgs boson is the last missing piece of puzzle for the SM. The confirmation of its existence will help to establish the SM as the 'true theory' of elementary particles.

One of the salient features of the SM is the neutral flavour-changing transition processes, the 'penguin' processes. Such processes do not occur at the tree level. It occurs, however, at the one-loop level which involves an internal W boson and a quark lines. The neutral change of flavour comes about because the coupling of W boson to the quark involves the CKM matrix, which mixes quarks from different generations. A penguin diagram therefore couples to quarks of different flavours, but of the same charge. The penguin diagram may be mediated by the gluon (gluon-penguin), the photon (photon-penguin), the $Z^0$ boson (Z-penguin), or the Higgs boson (Higgs-penguin) [3-22]. The Higgs-penguin differs from the other penguins in that the Higgs boson is scalar, whereas all others mediatory bosons are vector.

In this paper, we shall present the calculation of the Higgs-penguin, and obtain the on-shell vertex function for the Higgs-penguin. The result is then applied to processes of types: $H \to q_1 \bar{q}_2$ and $H \to M_1 M_2$.

## HIGGS-PENGUIN VERTEX FUNCTION

The calculation of Higgs-penguin is performed in the 'tHooft-Feynman gauge, which necessitates the introduction of the unphysical charged Higgs $\phi$ in addition to the W boson. Fig. 1 shows all the diagrams required in the calculation. Diagrams (a) to (d) are those with the Higgs boson couples to the W boson or the unphysical charged $\phi$. Diagrams (e) and (f) are with the Higgs boson couples directly to the quarks. The vertex function calculated from diagrams (a) to (f) is divergent, which can be eliminated by using a renormalization scheme proposed by Chia and Chong [16, 23]. This renormalization scheme introduces four one-particle reducible diagrams as shown in Fig. 1 (g) to (j). It is easily demonstrated that the counter term as calculated from the renormalization scheme above removes the divergence and yields a result identical to that would be obtained by using Ward-Takahashi Identity [24].

The proper vertex function is obtained by summing the contributions from all the diagrams in Fig. 1. Putting the external quark lines on shell, we arrive at the following expression for the on-shell vertex function:

$$\Gamma_{\text{on-shell}} = \frac{-g^3}{64\pi^2 M_W}(-m_2 L + m_1 R)\sum_j \lambda_j B_j \quad (1)$$

where $\lambda_j = V_{j2}^* V_{j1}$, $m_1$ and $m_2$ are the external quark masses, and $B_j$ is the Higgs-penguin vertex function given by

$$B_j = \frac{3}{2}\frac{m_j^2}{M_W^2} + \frac{m_j^4}{M_W^4}\frac{\ln(m_j^2/M_W^2)}{1 - m_j^2/M_W^2} + \int_0^1 dx \int_0^{1-x} dy \times$$

$$\times \{[-4y M_W^2 + (1-2y)m_j^2 + (2 - 2x - 3y + xy + y^2)k^2 + (1-y)k^2 m_j^2/M_W^2] D_W^{-1}$$

$$+ [2(2y-1)m_j^2 + y(y + x - 1)k^2 m_j^2/M_W^2 + 2y\, m_j^4/M_W^2] D_Q^{-1}$$

$$+ 2\ln(D_W/M_W^2) + 2(m_j^2/M_W^2)\ln(D_Q/M_W^2)\}, \quad (2)$$

and

$$D_W = x m_j^2 + (1-x)M_W^2 - y(1 + x - y)k^2, \quad (3)$$

$$D_Q = x M_W^2 + (1-x)m_j^2 - y(1 + x - y)k^2. \quad (4)$$

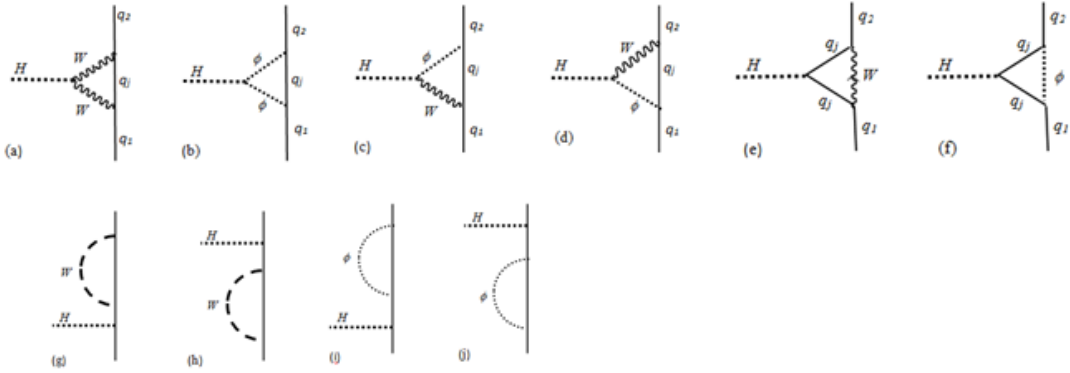

**FIGURE 1.** Fenyman diagrams employed in the calculation of Higgs-penguin vertex function. Diagrams (c) to (f) are diagrams with unphysical charged Higgs, and diagrams (g) to (j) are the one-particle reducible diagrams needed to remove the divergence.

In the expression in Eqs. (1) to (4), we have neglected the external quark masses in order to simplify the calculation. The salient feature of the vertex function does not change significantly by this approximation. As is obvious from Eqs. (3) and (4), the vertex function depends on the parameter $k^2$. Upon examining closely the integral in Eq. (1), it is easily shown that $B_j$ develops an imaginary part whenever $k^2 > (2m_j)^2$ or $(2M_W)^2$. As we are interested in the decay of Higgs boson, $k^2 = M_H^2$. At $M_H = 126$ GeV, $B_j$ takes on the following values:

$B_u = 7.702\text{E-}1 + i 8.532\text{E-}9$
$B_c = 7.706\text{E-}1 + i 8.722\text{E-}9$
$B_t = 9.588$ \quad (5)

It is noted that the integration in Eq. (1) at this value of $k^2$ does not cross the threshold for $M_W$ and $m_t$.

## DECAY RATE FOR $H \to q_1 \bar{q}_2$

We first consider the decay of Higgs boson into quark-antiquark pair of different flavours. To obtain the decay rate for $H \to q_1 \bar{q}_2$, we integrate over the phase space. This yields

$$\Gamma = \frac{1}{8\pi}\left(\frac{g^3}{64\pi^2}\right)^2 \frac{|\vec{q}_1|}{M_W^2 M_H^2}[m_1^2 m_2^2 + \frac{1}{4}(m_1^2 + m_2^2)(M_H^2 - m_1^2 - m_2^2)]\left|\sum_j \lambda_j B_j\right|^2 \quad (6)$$

where

$$|\vec{q_1}| = \frac{1}{2M_H}\sqrt{(M_H^2 - m_1^2 - m_2^2)^2 - 4m_1^2 m_2^2} \tag{7}$$

$$F_{kin} = \frac{|\vec{q_1}|}{M_W^2 M_H^2}[m_1^2 m_2^2 + \frac{1}{4}(m_1^2 + m_2^2)(M_H^2 - m_1^2 - m_2^2)] \tag{8}$$

$$F_{dyn} = \left|\sum_j \lambda_j B_j\right|^2 \tag{9}$$

The calculation of kinetic factor $F_{kin}$ is straightforward. To calculate the dynamic factor $F_{dyn}$, we need good estimates for the CKM matrix elements, including their relative phases. Fortunately these are available. The magnitudes of the CKM matrix elements are [25]

$$|\mathbb{V}| = \begin{pmatrix} 0.97425 & 0.2252 & 0.00393 \\ 0.230 & 0.953075 & 0.0406 \\ 0.0084 & 0.0387 & 0.9992 \end{pmatrix}. \tag{10}$$

The relative phases can be obtained by making use of the unitary relation $\sum_j \mathbb{V}_{ij}^* \mathbb{V}_{kj} = \delta_{ik}$. With the knowledge of the CKM matrix elements, the product of CKM matrix elements, $\lambda_j$, can be calculated. Its values are as shown in Table 1 for the different processes.

**TABLE 1.** Product of CKM matrix elements $\lambda_j$

| Process | $\lambda_u$ | $\lambda_c$ | $\lambda_t$ |
| --- | --- | --- | --- |
| $H \to s\bar{d}$ | 0.2140 | -0.2192 –i2.608E-4 | -1.940E-4 +i2.608E-4 |
| $H \to b\bar{d}$ | -1.666E-3 +i3.404E-3 | 9.338E-3 | -7.672E-3 –i3.404E-3 |
| $H \to b\bar{s}$ | 1.590E-5 +i8.759E-4 | -3.869E-2 –i8.759E-4 | 3.867E-2 |

Table 2 displays the values of the kinetic factor $F_{kin}$, the dynamic factor $F_{dyn}$ and the decay rate $\Gamma$ calculated for three processes considered.

**TABLE 2.** Table of $F_{kin}$, $F_{dyn}$ and the decay rate $\Gamma$ for the different processes

| Process | $F_{kin}$ | $F_{dyn}$ | $\Gamma$ ((GeV) |
| --- | --- | --- | --- |
| $H \to s\bar{d}$ | 2.205E-5 | 4.216E-4 | 7.185E-17 |
| $H \to b\bar{d}$ | 5.256E-2 | 4.576E-3 | 1.859E-12 |
| $H \to b\bar{s}$ | 5.258E-2 | 1.163E-1 | 4.725E-11 |

# DECAY RATE FOR $H \to M_1 M_2$

We next consider the decay of Higgs boson into a pair of pseudoscalar mesons with different flavours. To achieve this we have to connect the Higgs penguin vertex to the external meson states. Here we make the simplifying assumption that the pseudoscalar meson couples to quark-antiquark pair through a $\gamma_5$ coupling [26]. The process $H \to M_1 M_2$ therefore proceed through the loop diagram as shown in Fig.2. The loop diagram yields the following decay amplitude:

$$M = (-1)\int \frac{d^4 q_1}{(2\pi)^4} Tr(i\Gamma_{vertex}) \frac{i(\slashed{k} - \slashed{q}_1 - m_2)}{(k-q_1)^2 - m_2^2} g_2\gamma_5 \frac{i(\slashed{q}_1 - \slashed{P}_1 - m)}{(q_1 - P_1)^2 - m^2} g_1\gamma_5 \frac{i(\slashed{q}_1 - m_2)}{q_1^2 - m_2^2} \tag{11}$$

Here $k$ is the momentum of the $H$, $P_1$ is the momentum of $M_1$, and $g_1$ and $g_2$ are the coupling constants at the $M_1$-quark and the $M_2$-quark vertices respectively. The integration over the internal momentum $q_1$ is logarithmically divergent. We introduce a cut-off momentum $\Lambda$ to tame the divergence. Assuming $\Lambda$ to be large compared to all the masses involved in the integration gives

$$M = \left(\frac{-ig^3}{1024\pi^4 M_W}\right)(g_1 g_2 \ln\frac{\Lambda^2}{M^2})(m_1 - m_2)(m_1 + m_2 - m)\sum_j \lambda_j B_j \tag{12}$$

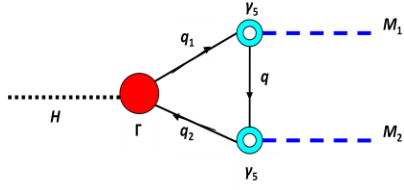

**FIGURE 2.** Diagram for the decay of Higgs boson into a pair of pseudoscalar mesons with different flavours.

The decay rate $\Gamma$ is then obtained by integrating over the phase space of $|M|^2$. After some algebra, the following expression for the decay rate is obtained:

$$\Gamma = \frac{1}{8\pi}\left(\frac{g^3}{1024\pi^4 M_W}\right)^2 \left(g_1 g_2 \ln\frac{\Lambda^2}{M^2}\right)^2 \frac{|\vec{P_1}|}{M_W^2 M_H^2}(m_1-m_2)^2(m_1+m_2-m)^2 \left|\sum_j \lambda_j B_j\right|^2 \quad (13)$$

In the above expression, the dynamic factor is the same as in Eq. (9). The kinetic factor, however, takes on a different expression:

$$F_{kin} = \frac{|\vec{P_1}|}{M_W^2 M_H^2}(m_1-m_2)^2(m_1+m_2-m)^2 \quad (14)$$

where

$$|\vec{P_1}| = \frac{1}{2M_H}\sqrt{(M_H^2-M_1^2-M_2^2)^2 - 4M_1^2 M_2^2} \quad (15)$$

The calculation is now applied to the following two decay modes of $H$:

(a) $H \to K^-\pi^+$

(b) $H \to B^-K^+$

The values of the parameter $g_1 g_2 \ln \Lambda^2/M^2$ in Eq. (13) are taken from the earlier fit to the rare decays of $K$ and $B$ mesons [26], as given in Table 3.

**TABLE 3.** Values of $g_1 g_2 \ln \Lambda^2/M^2$ as obtained from fit to the rare decays of $K$ and $B$ mesons

| Process | $g_1 g_2 \ln \Lambda^2/M^2$ |
|---|---|
| $K \to \pi e^+ e^-$ | 86.20 |
| $B \to K e^+ e^-$ | 85.44 |

Using these values of $g_1 g_2 \ln \Lambda^2/M^2$, the decay rates for the decay rates of the Higgs boson are as given in Table 4.

**TABLE 4.** Table of $F_{kin}$, $F_{dyn}$ and the decay rate $\Gamma$ for the different meson processes

| Process | $F_{kin}$ | $F_{dyn}$ | $\Gamma$ ((GeV) |
|---|---|---|---|
| $H \to K^-\pi^+$ | 4.750E-11 | 4.216E-4 | 4.611E-23 |
| $H \to B^-K^+$ | 2.861E-4 | 1.163E-1 | 7.527E-14 |

## CONCLUSION

We have calculated the Higgs-penguin vertex in the 'tHooft-Feynman gauge. The divergence is cured by utilizing a renormalization scheme developed earlier by Chia and Chong. The on-shell vertex function is obtained for the decay of Higgs boson into quark-antiquark pair of different flavours. $H \to q_1 \bar{q}_2$. The decay rate is found to be 7.185E-17 GeV, 1.859E-12 GeV and 4.725E-11 GeV respectively for the processes $H \to s\bar{d}$, $H \to b\bar{d}$ and $H \to b\bar{s}$.

The Higgs-penguin vertex is applied to the hadronic process of type $H \to M_1 M_2$. The pseudoscalar meson is assumed to couple to the quark through a $\gamma_5$ coupling. The resultant triangle diagram gives rise to a divergence which is tamed by introducing a cut-off momentum $\Lambda$. The couplings of quark to mesons $M_1$ and $M_2$ are denoted respectively by $g_1$ and $g_2$. The combinatory parameter $g_1 g_2 \ln \Lambda^2/M^2$ has been determined in an earlier fit to rare decays of $K$ and $B$ mesons. Using the values of $g_1 g_2 \ln \Lambda^2/M^2$ determined, the decay rate is found to be 4.611E-23 GeV and 7.527E-14 GeV respectively for the processes $H \to K^-\pi^+$ and $H \to B^-K^+$.

The decay rates for Higgs boson into pair of pseudoscalar mesons with different flavours are very small.

# REFERENCES

1. ATLAS Collaboration, *Phys. Lett. B* **716**, 1 (2012).
2. CMS Collaboration, *Phys. Lett. B* **716,** 30 (2012).
3. T. Inami and C.S. Lim, *Prog. Theor. Phys.* **65**, 297; *ibid* 1772E (1981).
4. A.J. Buras, *Phys. Rev. Lett.* **46**, 1354 (1981).
5. N.G. Deshpande and G. Eilam. *Phys. Rev.* **D26**, 2463 (1982).
6. N.G. Deshpande and M. Nazerimonfared, *Nucl. Phys.* **B213**, 390 (1983).
7. S.P. Chia and G. Rajagopal, *Phys. Lett.* **156B**, 405 (1985).
8. J.M. Soares and A. Barroso, *Phys. Rev.* D **39**, 1973 (1989).
9. E. Ma and A. Pramudita, *Phys. Rev.* D **22**, 214 (1980).
10. A. Axelrod, *Nucl. Phys.* **B209**, 349 (1982).
11. M. Clements *et al.*, *Phys. Rev.* D **27**, 570 (1983).
12. V. Ganapathi *et al.*, *Phys. Rev.* D **27**, 879 (1983).
13. G. Eilam, *Phys. Rev.* D **28**, 1202 (1983).
14. K.I. Hikasa, *Phys. Lett.* **148B**, 221 (1984).
15. M.J. Duncan, *Phys. Rev.* D **31**, 1139 (1985).
16. S.P. Chia and N.K. Chong, "The Z-penguin Vertex", in *Physics at the Frontiers of the Standard Model*, ed. N.V. Hieu and J.T.T. Van (Editions Frontieres, Gif-sur-Yvette, 1996) pp. 532-535.
17. S.P. Chia, N.K. Chong and T.L. Yoon, "The Electroweak Penguins from the Standard Model", in *Proceedings of the Inauguration Conference of the Asia-Pacific Center for Theoretical Physics: Current Topics in Physics*, ed. Y.M. Cho, J.B. Hong and C.N. Yang (World Scientific, Singapore, 1998), pp. 994-1003.
18. S.P. Chia, "The Absorptive Contribution of the $Z^o$-penguin", in *Proceedings of Seventh Asia Pacific Physics Conference*, ed. H.S. Chen (Science Press, Beijing, 1998), pp. 174-175.
19. S.P. Chia and W.L. Lim, "The Absorptive Contributions of the Electroweak Penguins", in *Proceedings of International Meeting on Frontiers of Physics 1998*, ed. S.P. Chia and D.A. Bradley (World Scientific, Singapore, 2000), pp. 361-371.
20. S.K. Krishnan, "Investigation of Neutral Change of Quark Flavours Involving Emission of Lepton-Antilepton Pair", M.Sc. Thesis, University of Malaya, 1998.
21. R.S. Willey and H.L. Yu, *Phys. Rev.* D **26**, 3086 (1982), *Phys. Rev.* D **26**, 3797 (1982).
22. B. Grzadkowski and P. Krawczyk, *Z. Phys.* **C18**, 43 (1983).
23. B.H. Tan, "Exact Calculation of the Flavour-Changing Quark-Higgs Vertex", M.Sc. Thesis, University of Malaya, 1997.
24. J.M. Soares and A. Barroso, *Phys. Rev.* D **39**, 19 (1989).
25. J. Beringer *et al*. (Particle Data Group), *Phys. Rev.* D **86**, 010001 (2012).
26. S.P. Chia, "Re-analysis of Rare Electroweak Decays of *K* and *B* Mesons", paper presented at *International Conference on Flavour Physics in the LHC Era*, Nanyang Technological University, Singapore, 8-12 November 2010.